\documentclass[aps,twocolumn,prb,amsmath,amssymb,showpacs,superscriptaddress]{revtex4}
\usepackage{graphicx,multirow,array}

\newcommand{\be}{\begin{equation}}
\newcommand{\ee}{\end{equation}}
\newcommand{\bea}{\begin{eqnarray}}
\newcommand{\eea}{\end{eqnarray}}

\newcommand{\ra}{\rangle}
\newcommand{\la}{\langle}

\def\tc{\mathrm{c}}

\def\ra{\rangle}
\def\la{\langle}

\begin{document}
\title{Rejuvenation and Memory in model Spin Glasses in 3 and 4 dimensions.}
\author{S.~Jim\'enez}
\affiliation{Dipartimento di Fisica, INFM and SMC, U. di Roma {\em La Sapienza},
P.le A. Moro 2, Roma I-00185, Italy.}
\affiliation{Instituto de Biocomputaci\'on y F\'{\i}sica de Sistemas
Complejos (BIFI), Corona de Arag\'on 42, Zaragoza 50009, Spain.}
\author{V. Mart\'{\i}n-Mayor}
\affiliation{Departamento de F\'{\i}sica Te\'orica I, Facultad de
Ciencias F\'{\i}sicas, Universidad Complutense, 28040 Madrid, Spain.}
\affiliation{Instituto de Biocomputaci\'on y F\'{\i}sica de Sistemas
Complejos (BIFI), Corona de Arag\'on 42, Zaragoza 50009, Spain.}
\author{S. P\'erez-Gaviro} \affiliation{Departamento de F\'{\i}sica
Te\'orica, Universidad de Zaragoza, Zaragoza 50009, Spain}
\affiliation{Instituto de Biocomputaci\'on y F\'{\i}sica de Sistemas
Complejos (BIFI), Corona de Arag\'on 42, Zaragoza 50009, Spain.}
\date{\today}
\begin{abstract}
We numerically study aging for the Edwards-Anderson Model in 3 and 4
dimensions using different temperature-change protocols.  In $D=3$,
time scales a thousand times larger than in previous work are reached
with the SUE machine. Deviations from cumulative aging are observed in
the non monotonic time behavior of the coherence length.  Memory and
rejuvenation effects are found in a temperature-cycle protocol,
revealed by vanishing effective waiting times.  Similar effects are
reported for the $D=3$ site-diluted ferromagnetic Ising model (without chaos).
However, rejuvenation is reduced if off-equilibrium corrections to the
fluctuation-dissipation theorem are considered.  Memory and
rejuvenation are quantitatively describable in terms of the growth
regime of the spin-glass coherence length.
\end{abstract}
\maketitle

\section{Introduction}

Nowadays the arena for comparisons between theory and experiments in
spin-glasses physics~\cite{BOOKS} is out-of equilibrium
dynamics.\cite{REV-DYN} Spin-glasses {\em age},\cite{AGINGDISCOVER} as
shown by the thermoremanent magnetization: consider a spin-glass that
has spent a time $t_w$ below its glass temperature, $T_\mathrm{c}$, in
the presence of a magnetic field. Let $t$ be the time elapsed since
the magnetic field was switched-off.  The magnetization decays as a
function of $t/t_w^\mu$, even for $t_w\sim 1$ day. The exponent $\mu$
could be 1 ({\em full-aging}),\cite{FULLAGING} although there are
recent experimental claims for $\mu$ being smaller than 1 ({\em
subaging}).\cite{CONTROFULLAGING} A somehow complementary experiment
consists in keeping the system for $t_w$ below its glass
temperature. Then, a magnetic field is switched on and the so-called
zero-field cooled (ZFC) magnetization is recorded while it
grows.\footnote{ The ZFC magnetization is a function of $t/t_w^\mu$,
$t$ being measured from the instant when the magnetic field was
switched on.} Full aging can be also observed in the a.c. magnetic
susceptibility, $\chi(\omega,t_w)$ which, for a fixed $\omega$,
decreases as $t_w$ grows. This time decay can be rescaled as a
$\omega$-independent function of $\omega t_w$,\cite{REV-DYN} although,
experimentally, one is restricted to the range $\omega t_w>1$.

{\em Memory} and {\em rejuvenation}~\cite{EXPDIP,EXP_REV} are sophisticated
manifestations of aging in experiments where the temperature is not kept
constant. Rejuvenation arises when changing temperature from $T_1$ to $T_2$
($T_1$ and $T_2$ smaller than the critical temperature, $T_\mathrm{c}$) a
system that has spent some time at $T_1$, so that $\chi(\omega,t_w)$ barely
depends on $t_w$.  Just after the $T_1\to T_2$ change, aging restarts. The
imaginary part of the susceptibility suddenly grows then relax.  The $t_w$
dependency of $\chi''(\omega,t_w)$ gets stronger, as for a {\em younger}
system. Rejuvenation means that the relaxation of $\chi''(\omega,t_w)$ is very
similar to the one of a system just quenched from $T>T_\mathrm{c}$ to $T_2$.
Sometimes it is said that the relaxation is identical to the one of a system
instantaneously quenched to $T_2$ from infinite temperature (in
Sect.~\ref{QUENCH-SECTION}, below, we elaborate on the different meaning of
{\em instantaneous temperature quench} in a experiment and in a computer
simulation).  If the susceptibility just after the quench to $T_2$ rises {\em
  above} the final value it had at $T_1$, one speaks of {\em strong
  rejuvenation}.\cite{TAKAYAMA} On the other hand, when the system is put back
at temperature $T_1$, $\chi''(\omega,t_w)$ continue its relaxation where it
left it just before the temperature change ({\em memory effect}). These
effects can also be observed in the real part of the susceptibility (see e.g.
Fig. 1 of Ref.~\onlinecite{MIYASHITA}), although rejuvenation is very
diminished as compared with the imaginary part. With the sophisticated {\em
  dip-experiment} temperature-change protocol,\cite{EXPDIP} memory and
rejuvenation are truly spectacular.

Memory and rejuvenation have been found in systems quite different
from spin-glasses (see, however,
Ref.~\onlinecite{DISSENTING}). Examples are structural
glasses~\cite{STRUCTGLASS}, polymers
(PMMA~\cite{PMMA-CILIBERTO,PMMA}), and systems not particularly glassy
(or not widely recognized as such), like colossal magnetoresistance
oxides.\cite{AGINGCMR} Moreover, a disordered ferromagnetic
alloy,\cite{DISFERR} becoming spin-glass at lower temperatures, has
shown rejuvenation and memory, through the dip-experiment protocol
(although in this case memory could be easily erased by lowering
the temperature). Spin-glasses display the quantitatively stronger
effects, but it is unlikely that the physical mechanism underlying
memory and rejuvenation are specific of spin-glasses.

The above definitions for memory and rejuvenation need
qualification. Under very small temperature changes~\cite{TWINS} (say
$\frac{T_1-T2}{T_1}< 5\times 10 ^{-3}$) the behavior of the spin-glass is
rather smooth. On the other hand, sharp memory and rejuvenation can be
observed~\cite{GHOST} for $\frac{T_1-T2}{T_1}\sim 0.07\,.$ The
crossover from small to drastic effects is rationalized using {\em
effective isothermal waiting times.}\cite{TWINS,COHE-LENGTH} Consider
the simplest temperature change protocol: a system is aged for time
$t_w$ at temperature $T_1$, then its temperature is suddenly shifted
from $T_1$ to $T_2$. After the shift, the ZFC magnetization is
measured.  The effective time, $t^\mathrm{eff,shift}_{T_2}$, is the
age of the isothermally aged system at temperature $T_2$, whose ZFC
magnetization\footnote{That is, one ages the system at $T_2$ for time
$t^\mathrm{eff,shift}_{T_2}$, then switches on the magnetic field and
records the growing magnetization.}  is most similar to the one of the
temperature-shifted system (the two relaxations are not
identical\cite{TWINS}).  Rejuvenation arises when
$t^\mathrm{eff,shift}_{T_2}/t_w$ is below experimental resolution.

Similarly, one can define~\cite{COHE-LENGTH} an effective time for the
temperature cycle protocol $T_1\rightarrow T_2\rightarrow
T_1$:\footnote{We somehow simplify the protocol description, for
details see Ref.~\onlinecite{COHE-LENGTH}.} one keeps the system a
time $t_w$ at $T_1$, then shifts the temperature to $T_2$, waits a
time $t_2\sim 20 t_w$, shifts back the temperature to $T_1$, switches
on a magnetic field and then records the ZFC magnetization. The
effective time $t^\mathrm{eff,cycle}_{T_1}$ is obtained by looking to
the system aged at temperature $T_1$ for a time
$t_w+t^\mathrm{eff,cycle}_{T_1}$ whose ZFC magnetization is most
similar to the one of the temperature-cycled system. One has memory,
as we defined it above, when $t^\mathrm{eff,cycle}_{T_1}/t_2$
gets below experimental resolution. A large variety of spin-glasses
experiments find\cite{COHE-LENGTH,CONTROFULLAGING} for  $T_1>T_2$
\begin{equation}
\frac{t^\mathrm{eff,cycle}_{T_1}}{t_w}=\mathrm{exp}\left[-\frac{T_1-T_2}{x_0
T_2}\right]\,,\label{CUMU_AGING_EXP_1}
\end{equation}
with\footnote{The actual value of $x_0$ depends both on $T_1$ and on
the anisotropy of the microscopic spin-spin interaction (the more
Heisenberg-like the interaction is, the smaller $x_0$
becomes~\cite{COHE-LENGTH}).}
\begin{equation}
x_0\sim 10^{-2}\,.\label{CUMU_AGING_EXP_2}
\end{equation}

The theoretical investigation of these phenomena is less advanced than
its experimental counterpart.  Memory and rejuvenation can be
recovered in the dynamics of abstract energy-landscape
models.\cite{ENERGYPICT} However, one wants to reproduce these
phenomena in the Langevin dynamics for the standard spin-glass model,
the Edwards-Anderson (EA) model.\cite{BOOKS} This dynamics for the EA
model can only be investigated by Monte Carlo simulation. Yet,
difficulties have arisen in numerical investigation of memory and
rejuvenation.\cite{TAKAYAMA,YOSHINO,RICCI,BERTHIER,ROMANOS}
Furthermore,  the progress achieved regards temperature-shift and
temperature-cycle experiments. The dip-experiment protocol remains
still as too complicated to be analyzed theoretically.

Experiments where $(T_2-T_1)/T_1$ is very small can be accounted for
by the {\em cumulative aging} scenario,\cite{TWINS,COHE-LENGTH}
consisting in the three following hypothesis:
\begin{enumerate}
\item Aging is ruled by the growth of a coherence
length,\cite{BERNARDI} signaling the building of a spin-glass order.
For isothermal aging, this length is named $\xi_{T}(t)$, $t$ being the
total time spent in the glass phase. This isothermal growth-law has
been studied in experiments~\cite{COHE-LENGTH} and
simulations,\cite{C4DEF,JAP-GROWTH} although the measured $\xi_{T}(t)$
grows by an small factor in both cases. Numerically, a power law
\begin{equation}
\xi_T(t)=A_T t^{z(T)},\quad z(T)=z_\mathrm{c}\frac{T}{T_\mathrm{c}}\,,
\label{XI-T-CONVERSION}
\end{equation}
fairly fits the data. However, more complicated rules have been
used.\cite{TWINS,COHE-LENGTH,GHOST,CONTROFULLAGING}

\item The coherence-length always grows with time. It behaves continuously upon
temperature changes.
\item Effective times follow from the {\em isothermal} growth of the
coherence length. Consider a temperature shift after aging for time
$t_w$ at $T_1$. One has
\begin{equation}
\xi_{T_1}(t_w)=\xi_{T_2}(t^{\mathrm{eff,shift}}_{T_2})\,.\label{CUMULATIVE-AGING}
\end{equation}
A time $t$ after the shift, the coherence length is
\begin{equation}
\xi^\mathrm{shift}(t)=\xi_{T_2}(t+t^\mathrm{eff,shift}_{T_2})\,.
\label{CUMULATIVE-AGING-2}
\end{equation}
Similar reasoning is used in the analysis of more
complicated temperature-change protocols.
\end{enumerate}
Eq.({\ref{CUMULATIVE-AGING}}) is used in an indirect way, both in the
analysis of simulations~\cite{ROMANOS,TAKAYAMA} and
experiments.\cite{TWINS,COHE-LENGTH,GHOST} Relations such as
(\ref{XI-T-CONVERSION}), obtained in a different experiment, are used
to convert the measured effective times into length-scales and
viceversa.  It is difficult to find in the literature {\em direct}
data on the behavior of the coherence length upon temperature
changes. A nice exception are the simulations of
Ref.~\onlinecite{YOSHINO} where Eq.(\ref{CUMULATIVE-AGING-2}) was
directly checked. Those simulation spanned $10^5$ Monte Carlo steps
(MCS). For comparison with experiments, recall that 1 MCS $\sim$ 1
picosecond.

Memory and rejuvenation appear as hardly compatible with the
cumulative aging. Experiments show\cite{TWINS} that
$\xi_{T_1}(t_w)<\xi_{T_2}(t^{\mathrm{eff,shift}}_{T_2})\,$ when the
measured effective times are converted in length-scales, both for
$T_2>T_1$ and $T_1<T_2$, in contradiction with
Eq.(\ref{CUMULATIVE-AGING}).

Two theoretical scenarios are currently being considered to account
for memory and rejuvenation. Rejuvenation was interpreted in terms of
temperature chaos,\cite{CHAOS} namely extreme sensitivity of {\em
equilibrium} states in the glass phase to small temperature
changes. An overlap-length, $l_0(T_1,T_2)$, is postulated to
exist. Features at scale smaller than $l_0$ are unaffected by a
temperature change $T_1 \to T_2$ while at larger scales the system is
completely reorganized.  Rejuvenation is then attributed to large
length scales and strong rejuvenation requires small $l_0$.  The
ghost-domain scenario (see Ref.~\onlinecite{GHOST} for a recent account)
allows to reproduce memory in the chaos scenario.  The other
scenario~\cite{BOUCHAUD} is closer in spirit to cumulative aging.
Rejuvenation after a negative temperature shift would arise from the
so-called fast modes involving length-scales smaller than
$\xi_{T_1}(t_w)$, that were equilibrated at $T_1$ but fall out of
equilibrium at $T_2$. Memory would arise from time and length scales
separation: back to temperature $T_1$, fast modes re-equilibrate very
fast so that aging continues from the previous $T_1$ state.

However, when it comes to actual calculations, it turns out that no
convincing memory and rejuvenation has been found in computer
simulations of $3D$ spin-glass models, either with a
two-temperatures~\cite{YOSHINO,BERTHIER,TAKAYAMA,ROMANOS} or with a
dip-experiment protocol.\cite{RICCI} When the behavior of the
coherence length is followed for times up to $10^5$ MCS,\cite{YOSHINO}
Eqs.(\ref{CUMULATIVE-AGING}) and (\ref{CUMULATIVE-AGING-2}) are
fulfilled even for $\frac{T_1-T_2}{T_2}\approx \pm 0.33\,.$
Consistently with this finding, when the temperature cycle protocol is
analyzed in the EA model,\cite{ROMANOS,RICCI-PRIVATE} the $x_0$ in
Eq.(\ref{CUMU_AGING_EXP_1}) turns out to be of order 1 rather than of
order $10^{-2}\,$. Should $x_0$ not decrease significantly for larger
times, the whole low-temperature phase of the EA model could be
accounted for by cumulative aging (i.e. the low-temperature phase
would not be a spin-glass phase).

This contradiction with experiments is puzzling. It could be
indicating that the EA model lacks some crucial
ingredient~\cite{RICCI} (maybe long-ranged dipolar interactions?).  Or
maybe memory and rejuvenation involve time and length scales
unaccessible to present-day simulations. Indeed, experiments are
performed on a time-scale which is about $10^8$ times longer than
typical simulations. Yet, experimentally,\cite{COHE-LENGTH} there are
around $\sim 10^5$ spins in a coherent cluster (hence $\xi_T(t_w)\sim
40$ lattice spacings), while simulations achieve (see below)
$\xi_T(t_w)\sim 10$ lattice spacings. When length scales are
confronted, the differences with experimental conditions do not seem
so dramatic.

As for higher space dimensions, a simulation~\cite{BERTHIER} of the
temperature cycle protocol for the 4D EA model, yielded strong
rejuvenation (as defined in Ref.~\onlinecite{TAKAYAMA}). Yet, results
in full agreement with cumulative aging, Eq.(\ref{CUMULATIVE-AGING}),
were reported for $\frac{T_1-T_2}{T_2}\approx \pm 0.125\,$ (the
simulation time was smaller than $10^4$ MCS). In the Migdal-Kadanof
lattice,\cite{SASAKI} where rather larger times can be simulated,
rejuvenation was found for $\frac{T_1-T_2}{T_1}\sim 0.1\,,$ suggesting
that $x_0$ in Eq.(\ref{CUMU_AGING_EXP_1}), does depends on the age of
the system.

In this work, we report simulations of a 3D (made with the SUE
machine~\cite{SUE}) and a 4D EA model with {\em binary} (rather than
{\em gaussian}~\cite{RICCI,YOSHINO,BERTHIER,SASAKI}) couplings. Our 3D
simulations are three orders of magnitude longer than previous ones.
We use real replicas~\cite{C4DEF} to study the coherence-length, that
is directly calculated (not inferred from effective times), through
the temperature changes.  In a temperature-cycle protocol, clear
memory and rejuvenation effects are found for large values of
$\frac{T_2-T_1}{T_1}\,$ both in 3D and in 4D
(sections~\ref{SECTION-STRONG} and~\ref{QUENCH-SECTION}). We also
observe strong rejuvenation (in the sense of
Ref.~\onlinecite{TAKAYAMA}), but only if we neglect corrections to the
Fluctuation-Dissipation theorem.\cite{FDT} The coherence-length is
shown to {\em decrease} upon some temperature changes, in
contradiction with cumulative aging. Moreover, a value
$t^\mathrm{eff,cycle}_{T_1}$ compatible with zero can be obtained for
$T_1=0.9 T_\mathrm{c}$ and $T_2=0.4 T_\mathrm{c}$, which is to be
expected in view of Eqs.(\ref{CUMU_AGING_EXP_1}) and
(\ref{CUMU_AGING_EXP_2}) (sections~\ref{QUENCH-SECTION}
and~\ref{XISECTION}).  Furthermore, we will show that the two-times
dependency of the time correlation function can be accounted for with
surprising accuracy by the coherence-length at the two relevant times,
both for isothermal aging and for temperature-shift protocols
(section~\ref{XITWOTIMES}). We perform exactly the same calculations
for the 3D {\em ferromagnetic} site-diluted Ising model~\cite{DILUTED}
(where, in the absence of frustration, chaos is absent), obtaining
very similar results. Although temperature chaos is probably present
in models for spin-glasses,\cite{SASAKI,RIZZO} our results in the
site-diluted {\em ferromagnetic} Ising model suggest that it plays no
role in producing memory and rejuvenation. This was maybe to be
expected, since memory and rejuvenation is being found experimentally
in materials where chaos in temperature seems to be absent or where a
thermodynamic glass transition has never been
found.\cite{STRUCTGLASS,PMMA-CILIBERTO,PMMA} Unfortunately, we have
made no progress\cite{SERGIOTESIS} in the analysis of the
dip-experiment protocol.

\section{Models and simulations}
Specifically, we consider Ising variables, $\sigma_i=\pm 1$, occupying
the nodes of a (hyper) cubic lattice in 3D and 4D with
nearest-neighbor, quenched disordered interactions. We report results
for the spin-glass with random $\pm 1$ couplings and the 3D site
diluted Ising ferromagnet~\cite{DILUTED} (spins are lacking with
probability $1-p$).
We evolve the system using  a sequential, local
heat-bath dynamics. Our time-unit (1 MCS $\sim$ 1 picosecond) is a
full-lattice update.  For the spin glass we studied the lattice size
$L=60$ in 3D (mostly in SUE), and $L=20$ for 4D (on PC clusters) with
some tests in $L=30$ finding no differences. For the site-diluted
model we studied $L=100$. The number of disorder realizations vary
within $16$ and $240$.

In the following, we will call a {\em direct-quench} to the procedure
of placing a fully disordered system (infinite temperature)
instantaneously at the working temperature.  This corresponds to an
infinite quenching-rate.

\begin{figure}[!hbt]
\includegraphics[width=\columnwidth ]{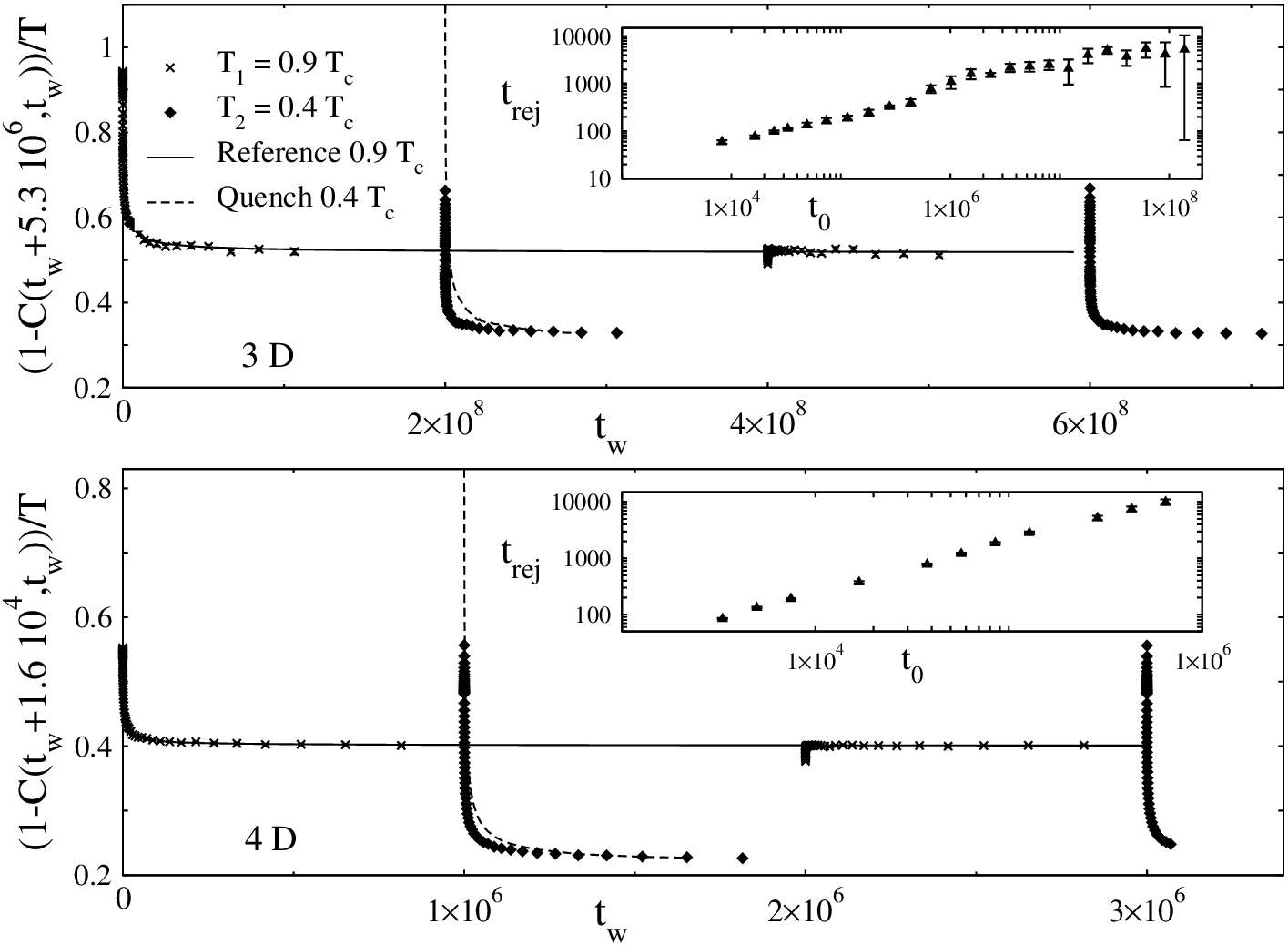}
\caption{{\bf Top:} Naive $\chi(\omega=2\pi/t_0,t_w)$, $t_0=5.3\times
10^6$ for the 3D EA model vs.  time. The $T$-cycle is $T_1\to T_2\to
T_1\to T_2$, each step lasting $t_s=2\times 10^8$.  The full line is a
reference run at $0.9 T_\mathrm{c}$. The inset shows the rejuvenation
time (see text) vs.  $t_0$. The dashed line is a direct-quench to
$T_2$. {\bf Bottom:} as top part for D=4, $t_0=1.6\times 10^4$ and
$t_s=10^6\,$.}
\label{FIG1}
\end{figure}

The Fluctuation-Dissipation Theorem (FDT) relates the autocorrelation
function in zero magnetic field \be C(t_w, t_w+t_0)=\frac{1}{V}
\sum_{i} \la \sigma_i(t_w) \sigma_i(t_w+t_0)\ra
\label{corrt}
\ee to the real part of the susceptibility:
$\chi(\omega=2\pi/t_0,t_w)\approx [1- C(t_w, t_w+t_0,)]/T\,$.  Yet,
off-equilibrium, FDT needs to be generalized replacing $T$ by $T/X[C]$
($X[C]$ is a smooth function~\cite{FDT} of $C(t_w, t_w+t_0)$).  Hence,
one assumes~\cite{YOSHINO,RICCI,BERTHIER,SASAKI} to be in
pseudo-equilibrium regime ($\omega t_w\gg 1\,$ thus $X[C]=1$), which
is not always true. We also obtain {\em spatial} information from the
correlation function of the overlap field,
$q_i(t)=\sigma_i^{(1)}(t)\sigma_i^{(2)}(t)$, built from two
independently evolving systems with the same couplings, at the same
temperature: \be C_4(r,t_w)=\frac{1}{V} \sum_{i} \la
q_i(t_w)q_{i+r}(t_w) \ra\,.
\label{corr4}
\ee

\section{Results}
\subsection{Strong rejuvenation?}\label{SECTION-STRONG}
In Fig.~\ref{FIG1} is shown the time-evolution of the naive
$\chi(\omega,t_w)$ ({\em i.e.} $[1- C(t_w, t_w+t_0)]/T\,$) for the EA
model in 3D (top) and 4D (bottom) for a (double) temperature cycle:
$T_1\to T_2\to T_1\to T_2$ ($T_1=0.9 T_\mathrm{c}$ , $T_2=0.4
T_\mathrm{c}$).  In 3D the system spends $t_s=2 \times 10^8$ MCS at
each temperature ($1000$ times longer than previous works), while in
4D $t_s=10^6$ MCS. The results of a reference run, with temperature
fixed to $T_1$, are also shown (continuous line).  When the
temperature drops to $T_2$, $\chi(\omega,t_w)$ increases over the
reference curve and starts a new relaxation ({\em strong
rejuvenation}~\cite{TAKAYAMA}). When temperature is back to $T_1$,
$\chi(\omega,t_w)$ catches the reference run almost instantaneously
({\em memory}).  We call $t_\mathrm{rej}$ to the time that the
rejuvenated $\chi(\omega,t_w)$ is above the reference run (see insets
in Fig.~\ref{FIG1}), that is found to grow consistently with $t_0$
(much faster in 4D). It is then conceivable that an effect of
macroscopic time-duration could be observed (experiments explore
$t_0\sim 10^{13}$ MCS). However, specially in $3D$,
$t_\mathrm{rej}<t_0$.  This implies that this strong rejuvenation is
confined to the regime $\omega t_w<1$, which is out of reach for
measurements of the a.c. susceptibility (note that strong rejuvenation is not
always observed experimentally in the real part of the
susceptibility~\cite{MIYASHITA}).

In agreement with Ref.~\onlinecite{BERTHIER}, the relaxing curve after the
temperature drop is independent of $t_s$ on the explored range
($t_s=10^6,2 \times 10^7$ and $2 \times 10^8$ MCS in $D=3$). Also
shown in Fig.~\ref{FIG1} is the relaxation of $\chi(\omega,t_w)$ for a
direct-quench to $T_2$ (dashed-line). Such an infinitely-fast
temperature drop is not realistic (see
section~\ref{QUENCH-SECTION}). Anyhow, the relaxation is not identical
to the one after the temperature shift, but the two become very
similar (in $D=3$, this happens for $t_w\sim 4 t_0$).  This is in
marked contrast with previous simulations where $t_s\sim 10^4$ and
$t_0=64$.\cite{TAKAYAMA} For such a short times, one needs $t_w\sim
500 t_0$ for the two relaxation curves to approach each other.


\begin{figure}[!h]
\includegraphics[width=\columnwidth ]{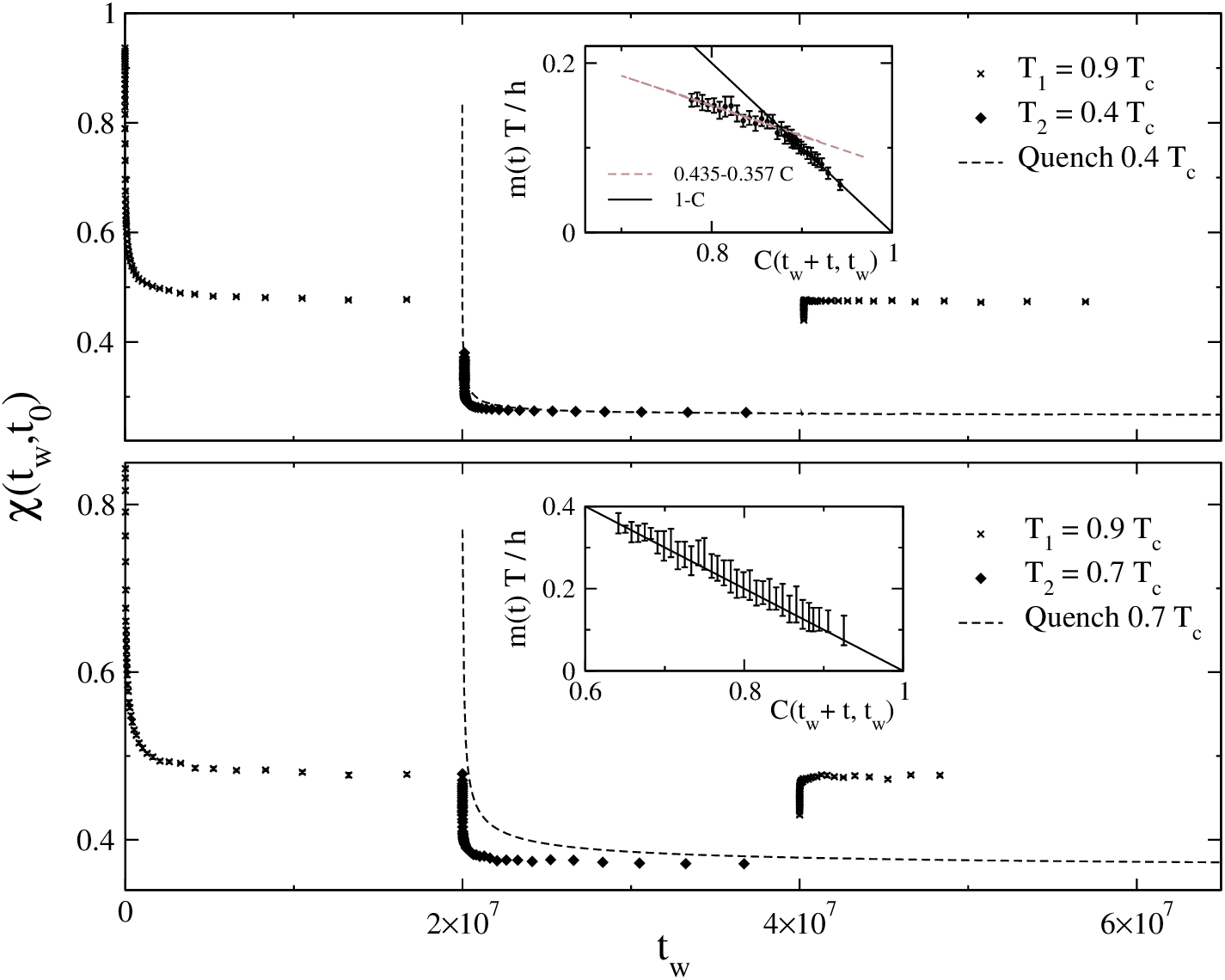}
\caption{{\bf Top:} As Fig.~\ref{FIG1} for $t_0=4.3\times 10^5$ and
$t_s=2\times10^7$ but correcting FDT violations.  {\em Inset:}
Off-equilibrium fluctuation dissipation relation (see text).  Data can
be nicely fitted to two straight lines.  {\bf Bottom:} As in top part
for $T_2=0.7 T_\mathrm{c}$.  Only the pseudo-equilibrium regime is
explored in this case (inset).}
\label{FIG2}
\end{figure}
Yet, in Fig.~\ref{FIG1} we assumed to be in pseudo-equilibrium
regime. In order to estimate the FDT correction factor $X[C]$ we use
the following procedure. We stay for time $t_s$ at $T_1$, then change
temperature to $T_2$, wait for time $t_w$ and switch-on a small
uniform magnetic field ($h=0.03$). We then record the magnetization,
$m(t_w+t)$, and $C(t_w, t_w+t)$.  The sought $X[C]$ factor is obtained
drawing $\chi T=m(t+t_w) T /h$ versus $C(t_w, t_w+t)$ (insets of Fig.
\ref{FIG2}). The resulting plot, $t_w$-independent for large
$t_w$,\cite{FDT} can be fitted with two straight-lines, yielding
$X[C]$. In Fig. \ref{FIG2} (top) we show the time evolution of the
$\chi(\omega,t_w)$ for the same cycle as Fig.~\ref{FIG1} with
$t_s=2\times10^ 7$.  Correcting with $X[C]$ reduces rejuvenation to
the point that {\em strong rejuvenation} is no longer seen. The
susceptibility no more grows at the temperature drop to $T_2$,
although the relaxation restarts and still collapses appreciably with
the $\chi(\omega,t_w)$ curve obtained from a direct-quench.  Similar
conclusions are drawn in 4D.\cite{SERGIOTESIS} We report in
Fig. \ref{FIG2} (bottom) results for a cycle with a smaller
temperature step ($T_1=0.9 T_{\tc},\, T_2=0.7 T_{\tc}$).  Here, (see
inset), we stay in pseudo-equilibrium regime and rejuvenation is {\em
stronger} for the smaller temperature drop, once the correcting $X[C]$
factor is considered.  However the collapse with the direct-quench
curve starts only for $t_w\sim 20 t_0$.

\subsubsection{The diluted ferromagnet}
\begin{figure}[!h]
\includegraphics[width=\columnwidth ]{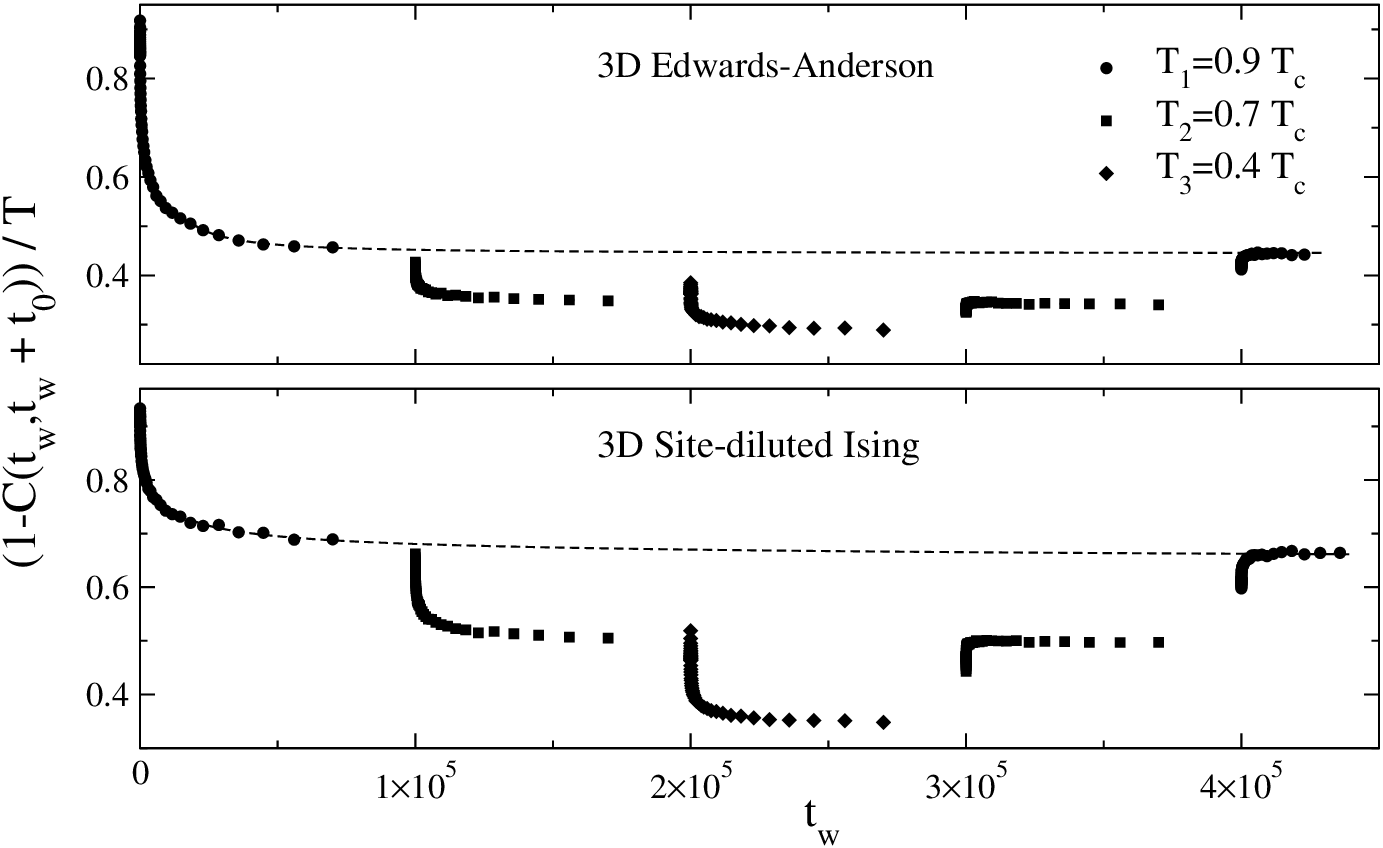}
\caption{Naive susceptibility in a 3 temperatures protocol (see text
for details), for the binary EA $D=3$ model ({\bf top}) and the $D=3$ site
diluted ferromagnetic Ising model ({\bf bottom}). The dashed line is a
direct-quench to $T_1$.}
\label{FIG4}
\end{figure}

Memory and rejuvenation have been found in other systems than
spin-glasses. A disordered ferromagnetic alloy,\cite{DISFERR} becoming
spin-glass at lower temperatures, has shown rejuvenation but much
weaker memory, through the dip-experiment protocol. For comparison, we
have simulated a site-diluted Ising model for $p=0.395$ ($T_{\tc}$ is
accurately known~\cite{DILUTED}).  All the interactions being
ferromagnetic, there is no temperature chaos in this system. We have
simulated a $L=100$ system checking that $\xi_T \ll L$ in our
simulation window. We measure the naive susceptibility with the
autocorrelation function (\ref{corrt}). Just for fun, we try a three
steps protocol, $T_1=0.9T_\mathrm{c}\to T_2=0.7T_\mathrm{c}\to
T_3=0.4T_\mathrm{c}\to T_2\to T_3$, staying $t_s=10^5$ MCS at each
temperature. The results are shown in Fig.~\ref{FIG4} (bottom)
together with the results for an equal protocol for the 3D EA model
(Fig.~\ref{FIG4}, top). In both cases rejuvenation and a {\em double}
memory are observed.  Also in two temperature
cycles~\cite{SERGIOTESIS} the susceptibility behaves as in the EA
model. To this level of analysis, there is no clear difference between
the Edwards-Anderson model and the site-diluted Ising model.

\subsection{Comparison with {\em experimental} direct-quench}\protect{\label{QUENCH-SECTION}}

\begin{figure}[!hbt]
\includegraphics[width=\columnwidth ]{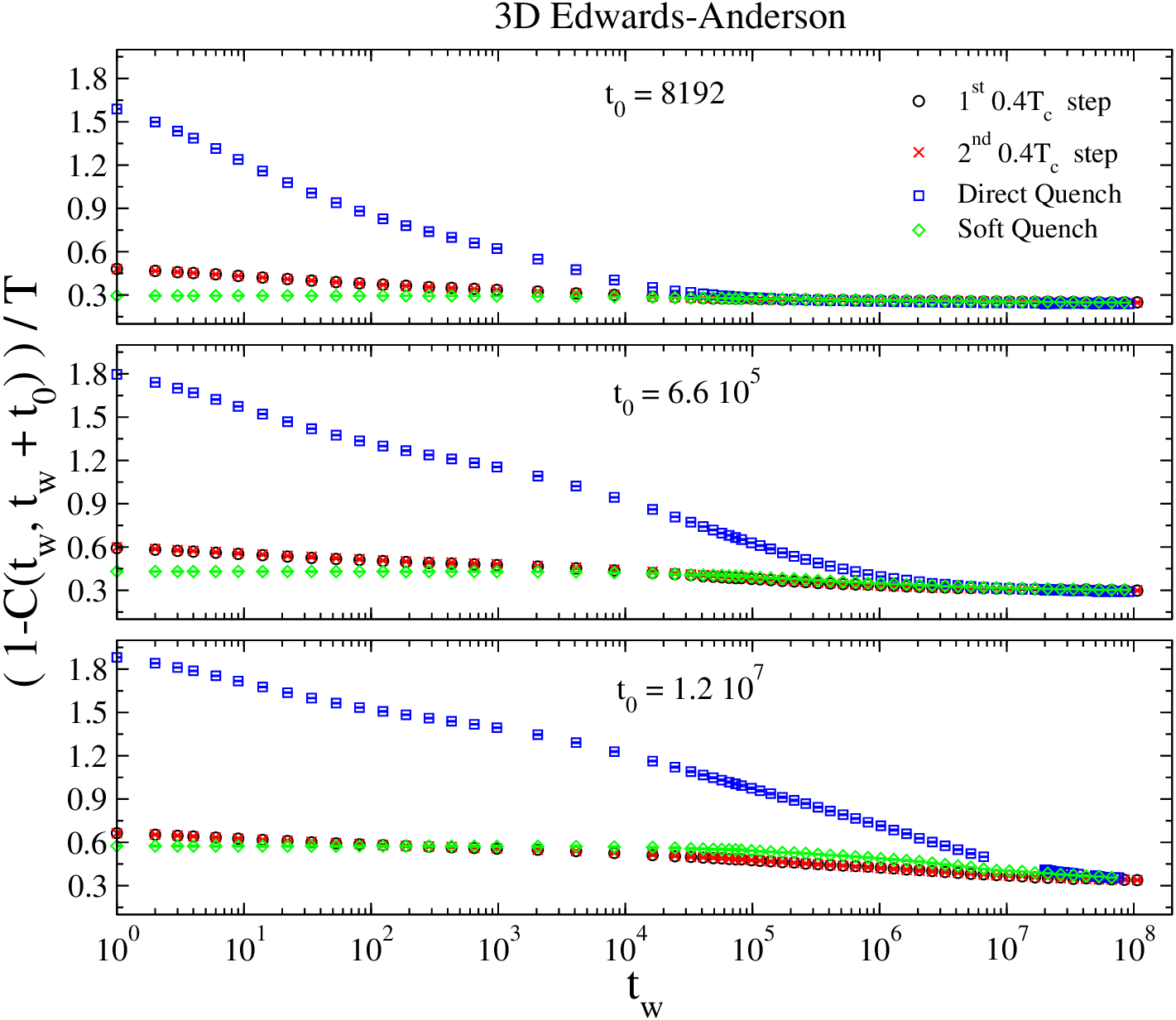}
\caption{(Color online) Naive $\chi(\omega=2\pi/t_0,t_w)$ of the 3D EA
model vs.  time, at temperature $T_2=0.4T_\mathrm{c}$, for several
values of $t_0$ and thermal histories.  In all cases, $t_w$ is
measured from the time of the (last) instantaneous quench to $T_2$.
{\bf Squares} correspond to the instantaneous drop from infinite
temperature to $T_2$. {\bf Circles} correspond to the system that has
spent $t_s=2\times 10^8$ MCS at $T_1=0.9T_\mathrm{c}$, then suffers an
instantaneous quench to $T_2$. {\bf Crosses} correspond to the system
that has been a time $t_s$ at $T_1$, then time $t_s$ at $T_2$, then
time $t_s$ at $T_1$ and finally suffers the instantaneous quench to
$T_2$. {\bf Diamonds} correspond to a gradual drop from
$9T_\mathrm{c}$ to $T_2$ in 20000 MCS (we incremented $1/T$ in
$0.113/T_\mathrm{c}$ every $10^3$ steps, the system spending
$1.2\times 10^4$ MCS in the spin-glass phase). }
\label{FIG2-I}
\end{figure}

In view of Eqs.(\ref{CUMU_AGING_EXP_1}) and (\ref{CUMU_AGING_EXP_2}),
and the large temperature drop that we are studying, one would expect
a perfect rejuvenation effect. However, Figs.~\ref{FIG1} and \ref{FIG2}
show that the relaxation after the first step at $0.9 T_\mathrm{c}$
considerably differs from the direct-quench (although this difference
is smaller than for shorter simulations~\cite{YOSHINO,TAKAYAMA}). This
seems in plain contradiction with experiments (see e.g. Fig.~4 of
Ref.~\onlinecite{GHOST}). Yet, upon reflection, one realizes that
the experimental direct-quench bears little resemblance with the
simulational one. In fact, the experimental sample that is
``instantaneously'' quenched to 0.4 $T_\mathrm{c}$, expends at least
10 seconds ($\sim 10^{13}$ MCS!) in the spin-glass phase.

In order to make a fair comparison with experiments, one should study
the relaxation after a ``soft'' quench (Fig.~\ref{FIG2-I}) from
high-temperature to the working temperature below the glass
transition. Yet, the fastest quenching rate that can be achieved in
experiments is far too slow to be reproduced in present-day computers.
To achieve a very slow temperature drop from high temperature to
working temperature, it is useful to consider Fig.~\ref{FIG1} in a
different way. One realizes that the system that has spent
$t_s=2\times 10^8$ MCS at 0.9$T_\mathrm{c}$, then suffers an
instantaneous temperature drop to 0.4$T_\mathrm{c}$ is a better
approximation to the experimental direct-quench to
0.4$T_\mathrm{c}\,$. In fact, the system spends quite a long time
close to the critical temperature, where the time evolution ---recall
Eq.(\ref{XI-T-CONVERSION})--- is faster.  When looking to the double
temperature cycle in Fig.~\ref{FIG1}, one needs to compare the
relaxation in the first and in the second steps at 0.4$T_\mathrm{c}$,
the first corresponding to the {\em reference} direct-quench, the
second being looked at as the {\em temperature-cycled} system.  This
comparison is shown in Fig.~\ref{FIG2-I}, together with the
relaxation after a {\em soft-quench}.

\begin{figure}[!bht]
\includegraphics[width=\columnwidth ]{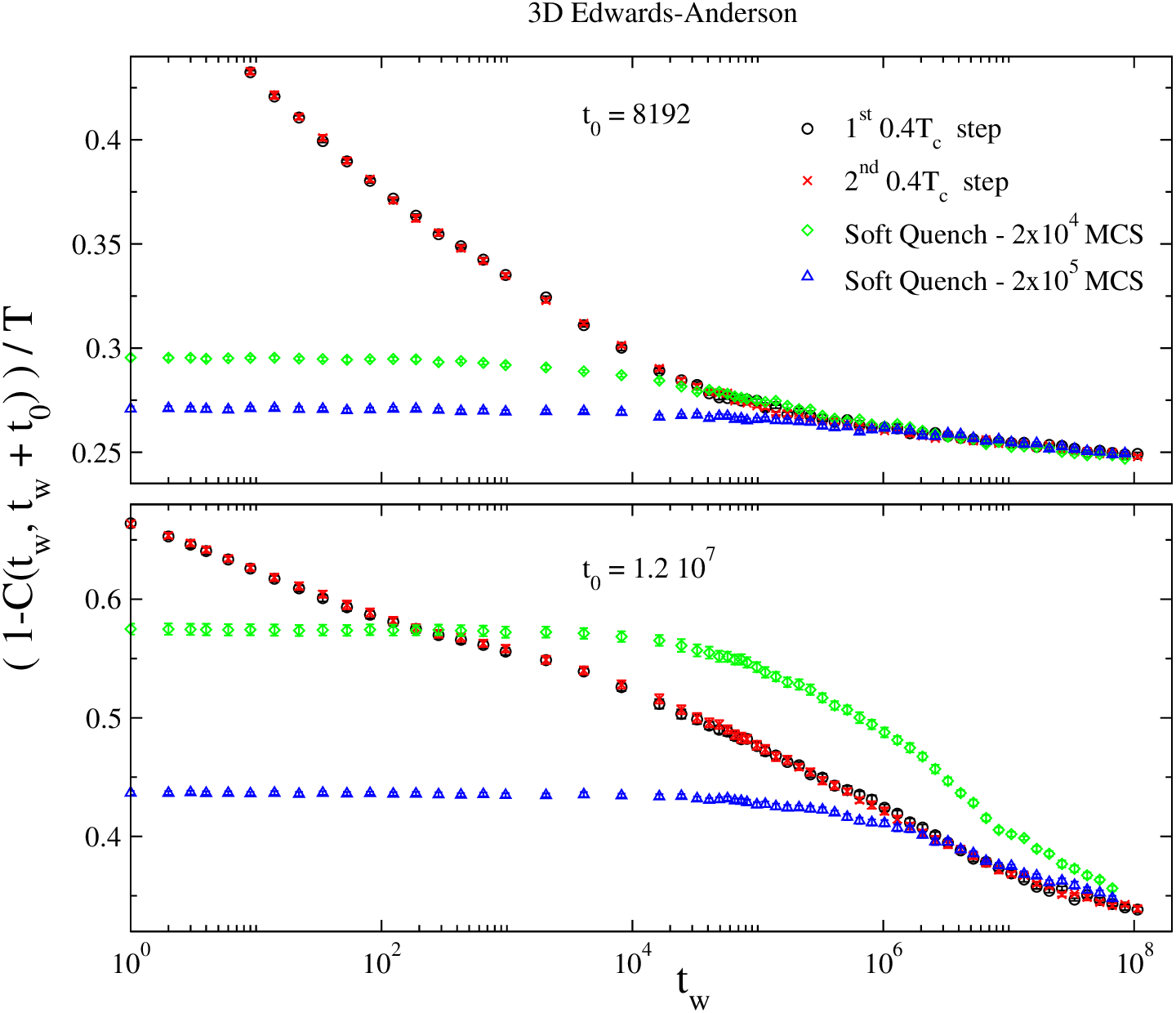}
\caption{(Color online) Close-up to the top and bottom panels of
Fig.~\ref{FIG2-I}, excluding the data for the instantaneous drop from
infinite temperature to $T_2=0.4T_\mathrm{c}$. We also include data
({\bf triangles}) for a still slower drop from $9T_\mathrm{c}$ to
$T_2$ in 200000 MCS (we incremented $1/T$ in $0.113/T_\mathrm{c}$
every $10^4$ steps, so that the system spends $1.2\times 10^5$ MCS in
the spin-glass phase).}
\label{FIG-SOFT}
\end{figure}

The frequencies shown in Fig.~\ref{FIG2-I} span three orders of
magnitude. In all cases, the relaxation for the softly-quenched
system,\footnote{Soft-quench in this context actually means not
infinite quenching rate, but dramatically faster than in experiments.}
that has spent $1.2\times 10^4$ MCS in the spin-glass phase, is much
closer to the one of the cycled-system than the one of infinite
quenching rate. Furthermore, the relaxations for the first and the
second steps at $0.4 T_\mathrm{c}$ are identical, up to our
statistical accuracy (see Fig.~\ref{FIG-SOFT}). This you may wish to
call {\em perfect rejuvenation}.

In Fig.~\ref{FIG-SOFT} we perform a detailed comparison between the
soft-quench (with two quenching rates) and the two-steps protocol. To
have a feeling of the frequency dependence, we show the smallest and
the largest frequencies in Fig.~\ref{FIG2-I}.  For very short times,
in the two-steps protocols we find a quick decay of the susceptibility
due to the sharp temperature drop. On the other hand, the
softly-quenched system shows a basically constant behavior (the slower
the quench, the lower the intial plateau is). When time becomes of the
order of the total time spent in the spin-glass phase during the
soft-quench, the susceptibility starts to decay and becomes very
similar to the two-steps protocol. At $t_0=8192$, the two
soft-quenches catch the relaxation of the two-step protocol and become
identical. At the smallest frequency, the fastest quench approaches
but does not catch the two-steps relaxation. On the other hand, for
the smallest quenching rate, the relaxation curve becomes identical to
the one of the two-steps protocol for $t_w \gtrsim t_0$, which
corresponds to the experimentally accessible time range.

\begin{figure}[!hbt]
\includegraphics[width=\columnwidth ]{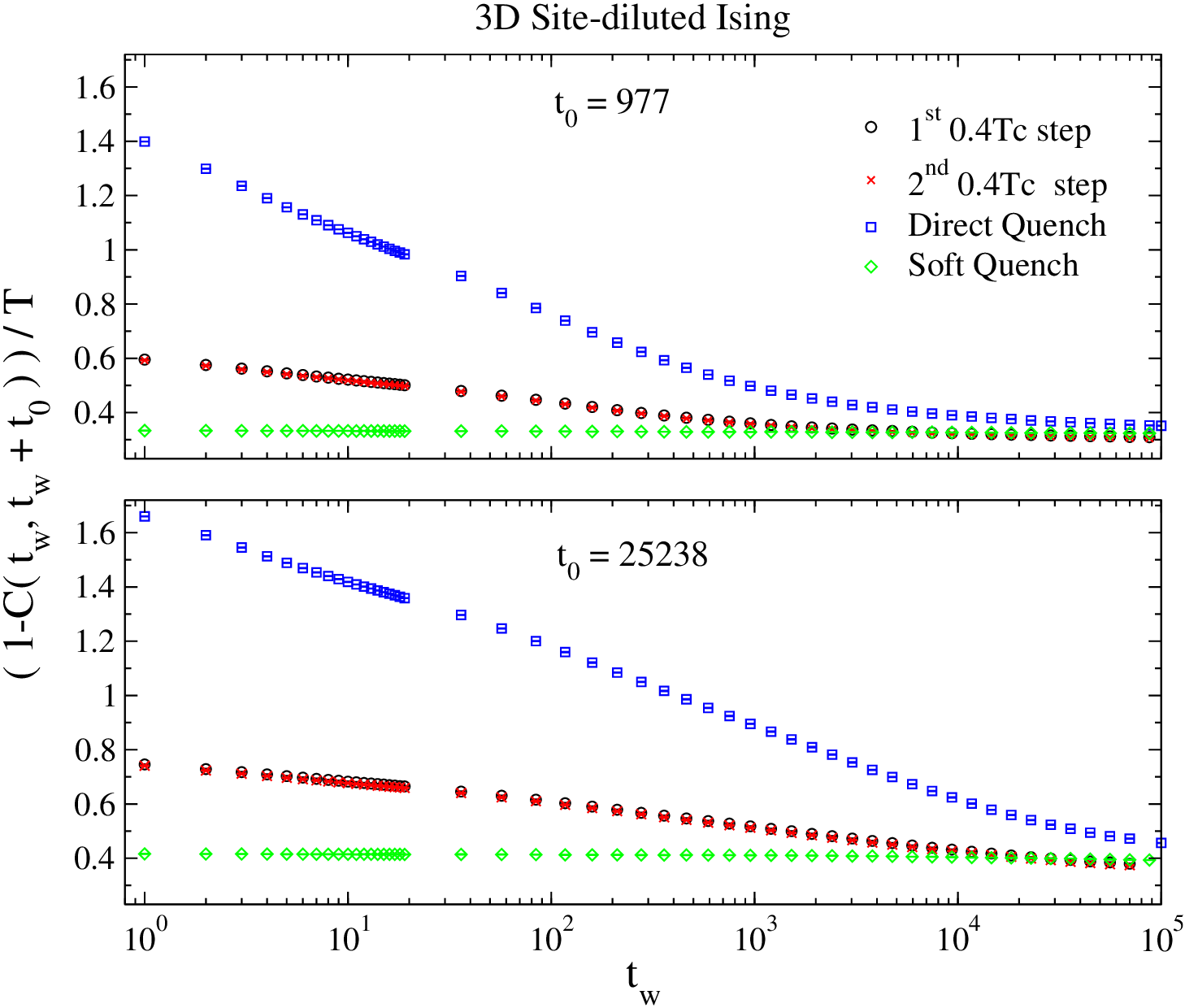}
\caption{(Color online) As in Fig.~\ref{FIG2-I}, for the diluted
ferromagnet and $t_s=10^5$ MCS. {\bf Diamonds} correspond to a gradual
drop from $9T_\mathrm{c}$ to $T_2$ in 20000 MCS (we incremented $1/T$
in $0.117/T_\mathrm{c}$ every $10^3$ steps). }
\label{FIG2-II}
\end{figure}

Quite similar results are obtained for the diluted ferromagnet, as we
show in Fig.~\ref{FIG2-II}. Although it cannot be noticed at the scale
of this plot, the relaxations in the first and second temperature-step
are not identical for the Ising model. They may be made to collapse if
the times in the second step are rescaled by a factor of $1.2$ (in a
protocol $0.7T_\mathrm{c}\rightarrow 0.4T_\mathrm{c} \rightarrow 0.7
T_\mathrm{c}\rightarrow 0.4T_\mathrm{c}$, the needed rescaling factor
is $2$).

\subsection{The coherence-length}\label{XISECTION}

The coherence-length may play a crucial role~\cite{BOUCHAUD} in this
physics, and should be followed in detail during temperature changes.
This was done previously in Ref.~\onlinecite{YOSHINO}, for times up to
$10^5$ MCS. Results in agreement with Eq.(\ref{CUMULATIVE-AGING-2})
were reported. We show here qualitatively different results for our
longer simulations in 3D.

The coherence-length may be obtained from non self-averaging integrals
of $C_4(r,t_w)$ using a second-momentum
estimator.\cite{COOPER,SUEFSS} Not having so many samples at our
disposal, we have obtained $C_4(r,t_w)$ (which {\em is} self-averaging 
for not very large $r$). The resulting curve has been 
fitted to~\cite{C4DEF} \be
C_4(r,t_w)=\frac{A}{r^\alpha} \mathrm{exp} \Big[ -\bigg(
\frac{r}{\xi(t_w)} \bigg)^\beta \, \Big]\,.
\label{ANSATZ}
\ee In 3D, we find fair fits in the range $2<r<20$, fixing
$\alpha=0.65$ and $\beta=1.7$ for all times and temperatures.  The
constant behavior of $\alpha$ does not agree with the results for the
4D model with Gaussian couplings.\cite{BERTHIER} To estimate errors in
the three parameters fit (\ref{ANSATZ}) is very difficult. To have a
feeling of their magnitude, let us report that $\alpha=0.7$ yields
good fits as well, with a $10\%$ increased $\xi$ estimate.

\begin{figure}[!hbt]
\includegraphics[width=\columnwidth ]{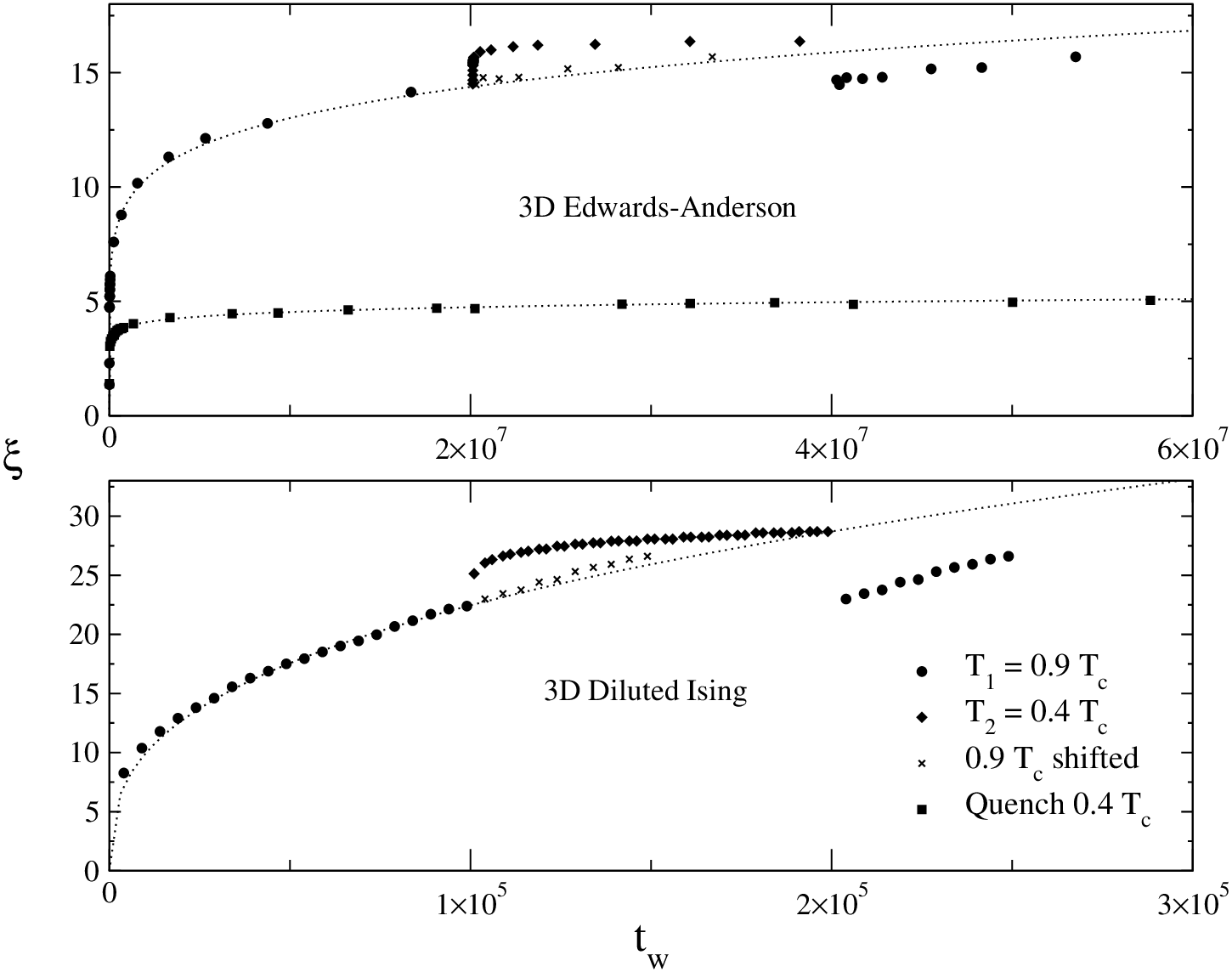}
\caption{ {\bf Coherence-length vs. time,} for the 3D EA model ({\bf
top}) and the diluted-ferromagnet ({\bf bottom}). The thermal history
has been a $T$-cycle $T_1 \to T_2 \to T_1,$ being $t_s=2 \times 10^7$
for the EA model and $t_s=10^5$ for the Ising-diluted model.({\em
Crosses:} second $T_1$-step data, translated back in time $t_s$ ; {\em
Dotted lines:} fits to $\xi(t_w)=At_w^x$ for $t_w< t_s$).}
\label{FIG3}
\end{figure}

See in Fig.~\ref{FIG3} (top), $\xi(t_w)$ for a direct-quench to
$T_2=0.4 T_\mathrm{c}$ and for a thermal cycle $T_1\to T_2\to T_1$
with $T_1=0.9 T_\mathrm{c}$ and $t_s=2\times 10^7$.  A power law with
exponent $\sim 0.144$ fits nicely $\xi_{T_1}(t_w)$ for $t_w< t_s$,
while the exponent for the direct-quench to $T_2$ is $\sim 0.065$
(full-lines in Fig.~\ref{FIG3}-top). Note that the exponent follow
Eq.(\ref{XI-T-CONVERSION}).  During the $T_2$-step, $\xi$ grows over
the $T_1$ value, and it is larger than for the direct-quench to $T_2$.
However, $\xi$ decreases when the system is back to $T_1\,$. Memory is
striking: data for the second $T_1$ step, if translated back $t_s$
MCS, are on top of the fit (obtained for $t_w<t_s\,$!).  Let us stress
two points regarding this result:
\begin{enumerate}
\item The coherence-length can {\em decrease} upon temperature
changes, violating cumulative-aging, Eq.(\ref{CUMULATIVE-AGING-2}),
and in contradiction with the time and length scales separation
scenario.\cite{BOUCHAUD,BOUCHAUDPRIVATE} However, the effect is not
symmetrical for negative and positive temperature shifts, as it was
inferred experimentally from effective-time measurements.\cite{TWINS}

\item The effective time for the temperature cycle is compatible with
zero (within our accuracy). This implies that, for $t_s\sim 10^7$,
$x_0$ in Eq.(\ref{CUMU_AGING_EXP_1}) is {\em not} of order one, as it
was found\cite{ROMANOS} for $t_s\sim 10^4\,$.
\end{enumerate}

As before (bottom part of Fig.~\ref{FIG3}), the behavior of the
diluted ferromagnet is completely analogous to the one of the EA model
(including the power-law growth of $\xi_T(t)$).

\subsection{The coherence length  and two times correlations}\label{XITWOTIMES}
\begin{figure}[!ht]
\includegraphics[width=\columnwidth ]{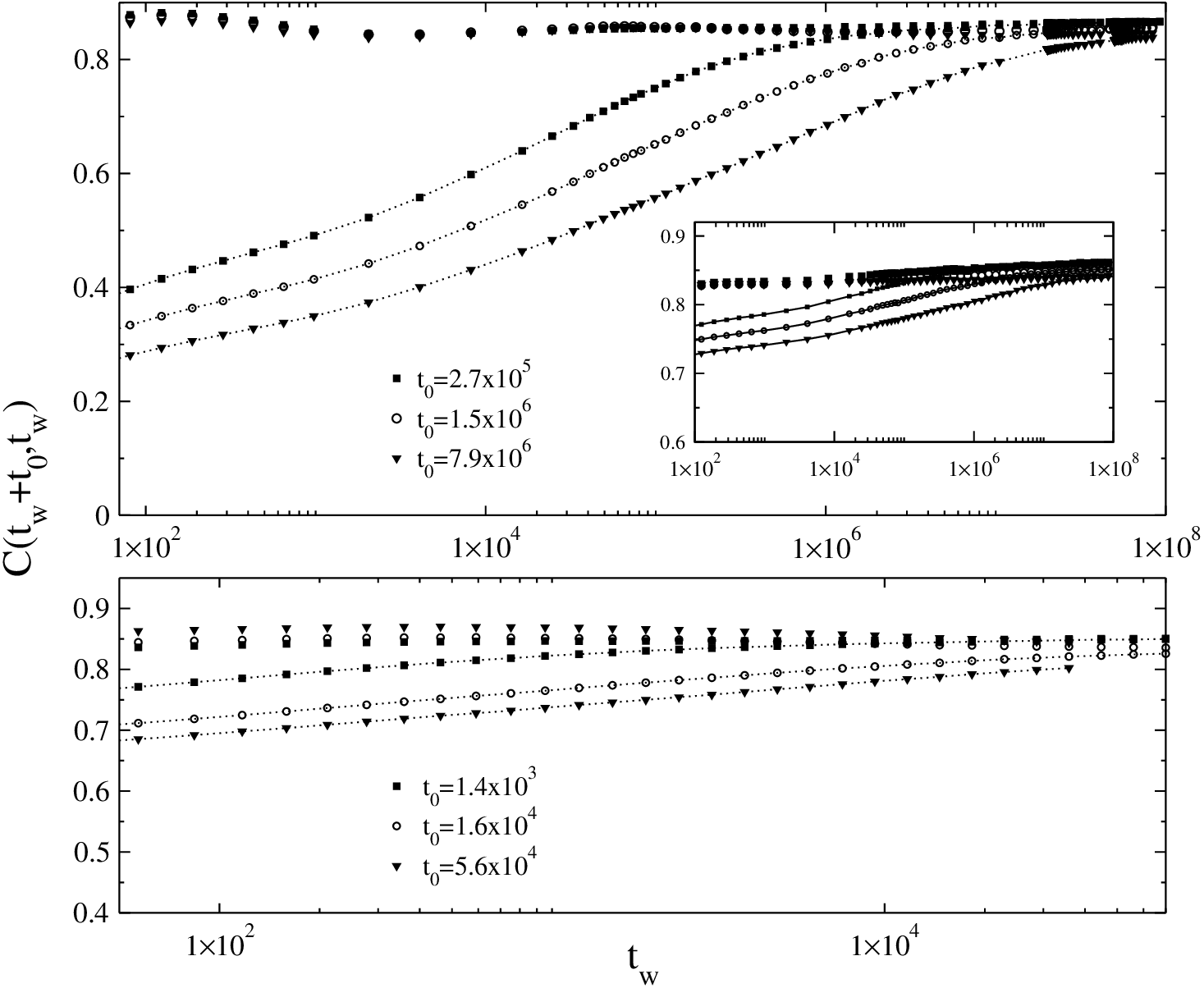}
\caption{{\bf Top}: $C(t_w, t_w+t_0)(\xi(t_0+t_w)/\xi(t_w))^{3/2}$
(points) and $C(t_w, t_w+t_0)$ (lines) vs. $t_w$, for the direct-quench
to $T_2$, for the EA model in 3D. {\em Inset:} As main plot for the
$T_2$ step of the T-cycle of Fig.\ref{FIG1}.  {\bf Bottom:} As inset of top panel for the diluted Ising model.}
\label{FIG3-II}
\end{figure}

A rather crucial feature of aging~\cite{REV-DYN} is that
two time-scales, $t_0$ and $t_w$, are involved.
One would like to relate the {\em one} time quantity $\xi_T(t_w)$, to
the {\em two} times correlation function. A crude estimate for $t_0\gg
t_w$ is 
\begin{equation}
C(t_w, t_w+t_0)\propto \frac{\xi^{D/2}(t_w)}{\xi^{D/2}(t_w+t_0)}\,,\label{VERYNAIVE}
\end{equation}
{\em i.e.} the coherent cluster that at time $t_w+t_0$ has linear size
$\xi(t_w+t_0)$, at time $t_w$ was composed of mutually incoherent
clusters of linear size $\xi(t_w)\,$. 

Indeed, (Fig.~\ref{FIG3-II}, top), the factor
$\xi^{3/2}(t_w+t_0)/\xi^{3/2}(t_w)$ absorbs almost all the $t_w$ and
$t_0$ dependency of $C(t_w, t_w+t_0)$, both for a direct-quench to
$T_2$ and for the $T_2$ part of the thermal cycle.  Note that even the
constant value for $C(t_w, t_w+t_0)\xi^{3/2}(t_w+t_0)/\xi^{3/2}(t_w)$
is equal for the direct-quench and for the thermal cycle.  Also at
$T_1$, $C(t_w, t_w+t_0)\xi^{3/2}(t_w+t_0)/\xi^{3/2}(t_w)$ is constant
within a band of width $5\%$ of its mean-value.\cite{SERGIOTESIS}
Quite similar results are obtained for the diluted ferromagnet, as we
show in the bottom part of Fig.~\ref{FIG3-II}. In spite of the
crudeness of the argument leading to Eq.(\ref{VERYNAIVE}) and the
uncertainty in the determination of $\xi$, the results are
surprisingly clear.

Note that if Eq.(\ref{VERYNAIVE}) was {\em exact} the dynamics would
be of the one-sector type,\cite{REV-DYN} which we do {\em not}
believe to be the case.\cite{SUEAGING} Anyhow, given
Eq.(\ref{VERYNAIVE}), full-aging~\cite{FULLAGING} is natural for a
power-law growth of $\xi(t_w)$.

Thus, memory and rejuvenation are driven by the rate-growth of
$\xi_T(t_w)$ rather than by its value or by the short-distance
behavior of $C_4(r,t_w)$.\cite{BOUCHAUD,BERTHIER} In our simulation,
rejuvenation is due to a {\em growth} of $\xi$ upon cooling (probably,
because of a sudden fall into a nearby energy minima), provoking a
change in the evolution of $C(t_w, t_w+t_0)$.  When temperature is
shifted back to $T_1$, $\xi_{T1}$ continues its growth as if it had
never being at $T_2$ with analogous consequences for the
correlation-function (memory). This implies a non monotonic behavior of
$\xi_T(t)$, in contradiction with cumulative-aging,
Eqs.(\ref{CUMULATIVE-AGING}) and (\ref{CUMULATIVE-AGING-2}).

\subsection{Results for $\omega t_w<1$}

Measurements of a.c. susceptibility are usually confined to the region
$\omega t_w >1$. On the other hand, thermal magnetoresistance
measurements can yield information on the time regime $t_0\gg t_w\,.$
It is therefore worthwhile to have a look to our correlation functions
in this regime.

The main issue here is the characterization of the time decay of
correlations (see Ref.~\onlinecite{SUEAGING} for a recent study).  One
finds~\cite{REV-DYN} that, at least when the correlation function lies
in some intervals (say $C_1< C(t_w,t_w+t_0) <C_2$), it behaves as a
function of $t_0/t_w^\mu$. It is possible that different intervals for
the correlation functions (usually called {\em
time-sectors}~\cite{REV-DYN}) are ruled by different exponents. The
existence of more than one time-sector is a necessary (but not
sufficient) requirement for dynamic ultrametricity.\cite{REV-DYN}
Some indications of the presence of more than one time-sector in the
dynamics of the EA model in 3D were found in
Ref.~\onlinecite{SUEAGING}.\\

\begin{figure}[!hbt]
\includegraphics[width=\columnwidth ]{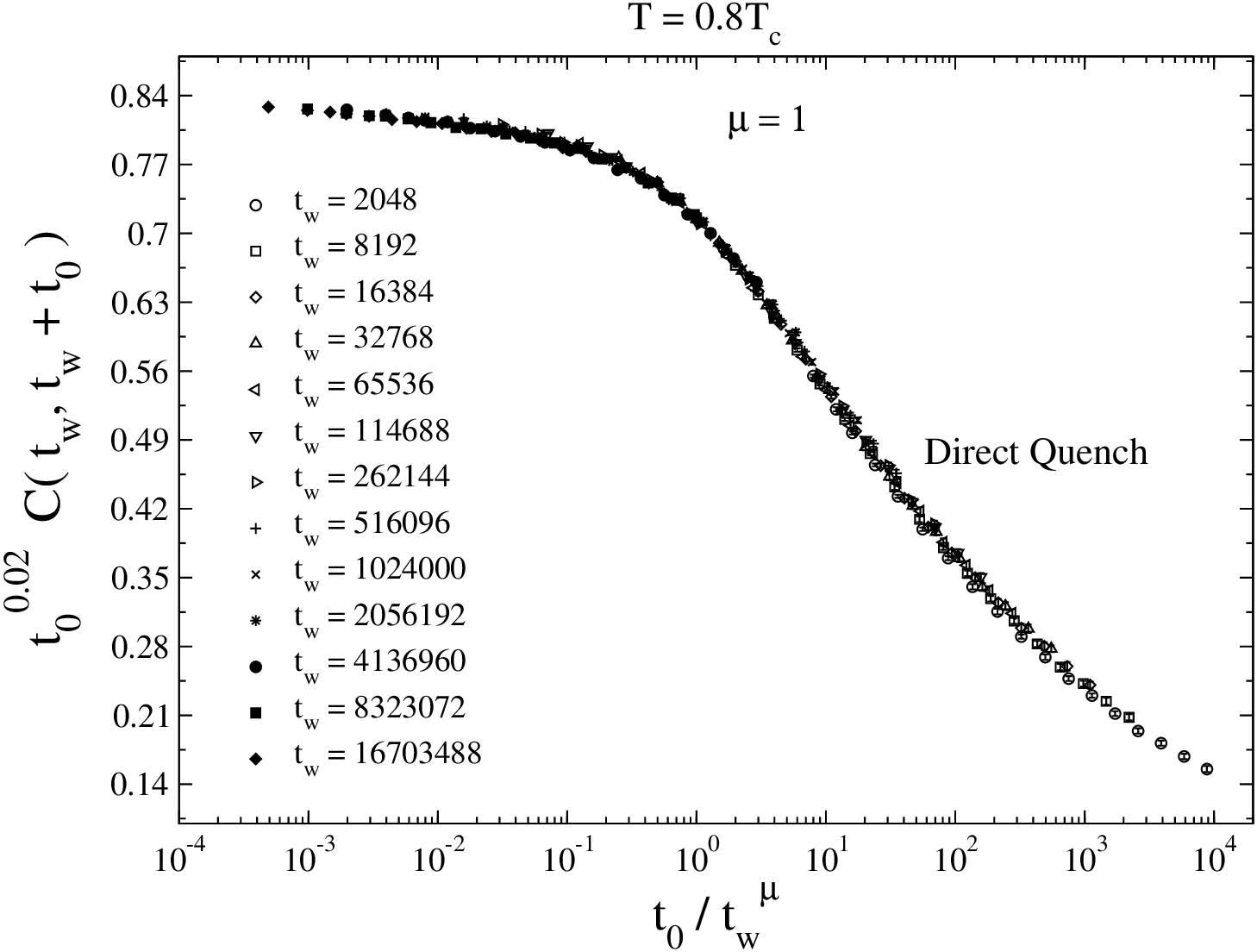}
\caption{Correlation function $C(t_w,t_w+t_0)$ scaled by $t_0^{0.02}$
(see Eq.(\ref{EQRIEGER})) after a direct-quench to $T=0.8
T_\mathrm{c}$, versus $t_0/t_w.$}
\label{FIG5-I}
\end{figure}

The experimental value of $\mu$ is
controversial. Recently,\cite{FULLAGING} it was claimed that it is
$\mu=1$ (full-aging). Previous failures in recognizing this, were
ascribed to quenching-rate effects~\cite{FULLAGING} (recall also
Sect.~\ref{QUENCH-SECTION}). In the quickest possible quench, $\mu=1$
was clearly identified. This interpretation was recently disputed 
by the Saclay group,\cite{CONTROFULLAGING} that find $\mu<1\,$.

Regarding computer simulations, the following scaling form for  the
time correlation function
was proposed,\cite{RIEGER}  and found to work  for a restricted $t_0$
and $t_w$ range: 
\begin{equation}
C(t_w,t_w+t_0)= t_0^{-x(T)} \Phi\left(\frac{t_0}{t_w}\right)\,.
\label{EQRIEGER}
\end{equation}
This equation implies that full aging should be observed in the
$t_0\gg t_w$ regime. More recent and longer simulations~\cite{SUEAGING}
found deviations from Eq.(\ref{EQRIEGER}), at least at some temperatures.

\begin{figure}[!htb]
\includegraphics[width=\columnwidth ]{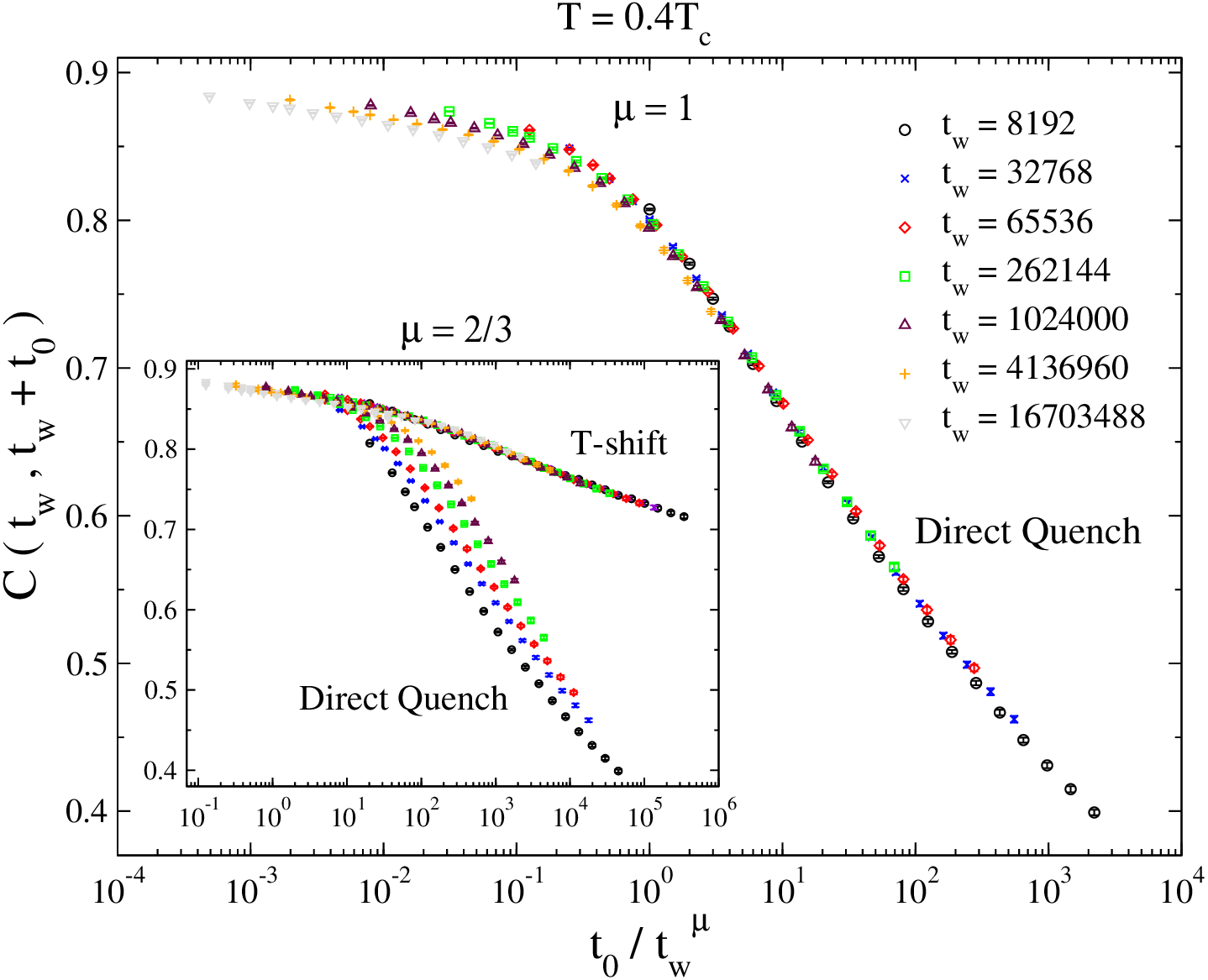}
\caption{(Color online) Correlation function $C(t_w,t_w+t_0)$ after a direct-quench
to $T=0.4 T_\mathrm{c}$, versus $t_0/t_w.$ {\bf Inset:} Data of main
plot, versus $t/t_w^\mu$, $\mu=2/3$. We also plot data for a system
that has been $2\times 10^8$ MCS at 0.9$T_\mathrm{c}$, then suffers a
temperature shift to $0.4 T_\mathrm{c}\,$.  }
\label{FIG5-II}
\end{figure}

In Fig.~\ref{FIG5-I} we plot our data for a direct-quench from
infinite temperature to $0.8T_\mathrm{c}$.  Eq.(\ref{EQRIEGER}) works
nicely with $x(T)=0.02$, which can be interpreted as evidence for
full-aging behavior.  However, at $T=0.4T_\mathrm{c}$ (see
Fig.~\ref{FIG5-II}), we have being unable of finding a working
$x(T)\,$. Actually, the relaxation is better interpreted in terms of
two time-sectors: for $t_0\gg t_w$, $\mu=1$ seems to provide a proper
scaling, while, at shorter $t_0$, $\mu=2/3$ does a better job.

In the inset of Fig.~\ref{FIG5-II}, we compare the relaxations at
$T=0.4T_\mathrm{c}$ for the direct-quench and for the system that
stayed $2\times 10^8$ MCS at $0.9T_\mathrm{c}$, then suffers a
temperature shift to $0.4T_\mathrm{c}$ (recall that this is our
slowest quenching-rate from infinite temperature). For the slowly
quenched system, the relaxation belongs to the $\mu=2/3$
time-sector. Moreover, the relaxation after the $T$-shift provides a
limiting curve, in the large $t_w$ limit, for the $\mu=2/3$ piece of
the direct-quench relaxation.

At least within the time range that we can study, it seems that the
quenching-rate does exert influence in the measured value of $\mu$ (as
proposed in Ref.~\onlinecite{FULLAGING}).

\section{Conclusions}

In this work, we have compared memory and rejuvenation effects for the
$D=3$ and $D=4$ binary EA model and for the $D=3$ the site-diluted
Ising model, finding quite similar results. Maybe the most important
question we have addresed is whether the Edwards-Anderson model has
a spin-glass low temperature phase, or not. The longer time scale
reachable with the SUE machine~\cite{SUE} (three orders of magnitude
longer than previous work), has allowed us to conclude that
spin-glass behave like experimental spin-glasses in a number of ways:
\begin{enumerate}
\item The relaxation of the a.c. susceptibility after a large
temperature shift is as for a {\em direct-quench}, provide that one
does not interpret the term {\em direct-quench} literally in the
simulations (Sect.\ref{QUENCH-SECTION}). We find little dependency in
the previous thermal history, once microscopically short times are
neglected.

\item Values of $x_0$, defined in Eq.(\ref{CUMU_AGING_EXP_1}), are not
of order 1 for waiting times $\sim 10^7$ MCS (contrary to the finding
of Ref.~\onlinecite{ROMANOS} for waiting times of order $\sim 10^4$
MCS).  This time dependence of $x_0$ is consistent with the findings
in the Migdal-Kadanof lattice.\cite{SASAKI}

\item The coherence length may {\em decrease} upon temperature
changes, in disagreement with the cumulative aging scenario, and in
agreement with recent experiments.\cite{TWINS,GHOST}

\item The growth-rate of the coherence length rules the decay of the
time-correlation function.\cite{BERNARDI} A very simple formula,
Eq.(\ref{VERYNAIVE}), accounts for the behavior of $C(t_w,t_w+t_0)$
with surprising accuracy, for different temperature protocols.
Contrary to the findings of short 4D simulations,\cite{BERTHIER} we do
not find a strong temperature dependency of the replica-field
correlation function, Eq.(\ref{corr4}), at short distances. We rather
ascribe rejuvenation to the coherence-length, which rules the
long-distance behavior of the correlation-function.

\item The exponent $\mu$ for the $t_0/t_w^\mu$ scaling of the decay of
$C(t_w,t_w+t_0)\,$, depends on the quenching-rate. In agreement with
recent experiments,\cite{FULLAGING} we find that the faster the quench
is, the easier it becomes to obtain $\mu=1$ (see, however,
Ref.~\onlinecite{CONTROFULLAGING} for experiments contradicting this
expectation).
\end{enumerate}

The above findings suggest that the 3D EA model behaves as
experimental spin-glasses do, contrary to what was inferred from
previous shorter simulations.\cite{RICCI,YOSHINO,TAKAYAMA,ROMANOS}
However, there are a few important differences.  {\em First,} the
out-of-phase and the in-phase susceptibility behaves quite differently
in experimental spin-glasses (see
e.g. Ref~\onlinecite{MIYASHITA}). This seems not to be the case for
the EA model,~\cite{TAKAYAMA,ROMANOS,SERGIOTESIS} within the
accessible time-window. {\em Second,} the coherence-length is not
found to decrease under positive temperature shifts, in contradiction
with the inferred behavior from experimental measurements of
effective-times.~\cite{TWINS,GHOST} {\em Third,} in the dip-experiment
protocol we have found~\cite{SERGIOTESIS} no memory and very weak
rejuvenation, in agreement with Ref.~\onlinecite{RICCI} even if we
study frequencies 15 times smaller. {\em Fourth,} recent
simulations do not~\cite{ROMANOS,BERTHIERYOUNG} find stronger memory
and rejuvenation effects for Heisenberg than for Ising models of
spin-glasses. This finding is in contradiction both with experiments
(see e.g.~Ref.\onlinecite{COHE-LENGTH}) and with a recent calculation of
overlap lengths in the Migdal-Kadannoff approximation.\cite{FLOREN}
The physical reasons underlying these differences between our best
model for spin-glasses and experimental spin-glasses are not yet
understood. It is of course possible that longer times need to be
studied. However, when time scales are converted to length-scales, one
finds that numerical coherence-lengths are smaller than the
experimental ones only by a small factor (see
also Ref.~\onlinecite{HUKUSHIMA-IBA}).

We have shown that the ferromagnetic site-diluted Ising model (where
chaos is absent) follows very closely the behavior of the EA model, at
least in the $\omega t_w >1$ regime. This is not totally unexpected,
as we already know experimentally that systems quite different from
spin-glasses show memory and
rejuvenation.\cite{STRUCTGLASS,DISFERR,PMMA-CILIBERTO,PMMA,AGINGCMR}
The natural conclusion is that chaos, although probably present in
realistic spin-glasses,\cite{SASAKI,RIZZO,FLOREN} need not be invoked
to explain memory and rejuvenation. However, it has been
argued~\cite{DISSENTING} that the experimental protocol introduced in
Ref.~\onlinecite{TWINS} may help to discriminate temperature chaos
from a somehow trivial restarting of the dynamics (yet, see
Ref.~\onlinecite{BERTHIER2003}). More work will be needed to assess
the usefulness of this new classification~\cite{DISSENTING} of aging systems.

\section{Acknowledgments}
We thank G. Parisi, A. Taranc\'on, L.A. Fern\'andez,
F. Ricci-Tersenghi, E. Marinari, and A.  Maiorano for discussions. 
Numerical simulations have been carried out in the dedicated computer
SUE and in the PC clusters {\em RTN3} and {\em RTN4}, of U. de
Zaragoza. S.J. and S.P.-G. acknowledge finantial support from DGA
(Spain) and (S.J.)  from the ECHP programme, contract
HPRN-CT-2002-00307, DYGLAGEMEM.  This work has been financially
supported by MEC (Spain) though research contracts FPA2000-0956,
FPA2001-1813, BFM2003-08532 and FIS2004-05073-C04.



\begin{thebibliography}{99}

\bibitem{BOOKS} See e.g. {\em Spin Glasses and Random Fields}, Ed.
A. P. Young. World Scientific (Singapore, 1997).
\bibitem{REV-DYN} J.P. Bouchaud, L. Cugliandolo, J. Kurchan and
M. M\'ezard, in Ref.~\onlinecite{BOOKS}; J.J. Ruiz-Lorenzo, {\em Advances in
Condensed Matter and Statistical Mechanics}, Ed. E. Korutcheva,
R. Cuerno, Nova Science Publishers (2004).
\bibitem{AGINGDISCOVER} R.V. Chamberlin, M. Hardiman and R. Orbach,
J. Appl. Phys. {\bf 52,} 1771 (1983); L. Lundgren, P. Svedlindh,
P. Nordblad, and O. Beckman, Phys. Rev. Lett. {\bf 51,} 911 (1983) and
J. Appl. Phys. {\bf 57,} 3371 (1985).
\bibitem{FULLAGING} At least for $10^{-2}< t/t_w< 10^2$, 
G.F. Rodriguez, G.G. Kenning, R. Orbach, Phys. Rev. Lett. {\bf 91,}
037203 (2003).
\bibitem{CONTROFULLAGING} V. Dupuis, F. Bert, J.-P. Bouchaud,
J. Hammann, F. Ladieu, D. Parker and E. Vincent, cond-mat/0406721.
\bibitem{EXPDIP} K. Jonason, K. Jonason, E. Vincent, J. Hammann,
J. P. Bouchaud, and P. Nordblad, Phys. Rev. Lett. {\bf 81,} 3243
(1998).
\bibitem{EXP_REV} L. Lundgren, P. Svendlinsh, O. Beckman, Journal of
 Magn. Magn. Mat. {\bf 31-34,} 1349 (1983); T. Jonsson, K. Jonason,
 P. J\"onsson, and P. Nordblad, Phys. Rev. B {\bf 59,} 8770(1999);
 J.Hammann, E.Vincent, V. Dupuis, M. Alba, M. Ocio and J.-P. Bouchaud,
 J. Phys. Soc. Jpn. {\bf 69,} (2000) Suppl. A, 206-211.
\bibitem{TAKAYAMA} H. Takayama and K. Hukushima,
J. Phys. Soc. Jpn. {\bf 71,} 3003 (2002).
\bibitem{MIYASHITA} S. Miyashita, E. Vincent, Eur. Phys. J. B {\bf
22,} 203 (2001).
\bibitem{DISSENTING} P.E. J\"onsson, H. Yoshino, H. Mamiya and
H. Takayama, Phys. Rev. B {\bf 71,} 104404 (2005).
\bibitem{STRUCTGLASS}
H. Yardimci, R.L. Leheny, Europhys. Lett. {\bf 62,} 203 (2003).
\bibitem{DISFERR} E.Vincent,V. Dupuis, M. Alba,
J. Hammann,J.-P. Bouchaud, Europhys. Lett {\bf50,} 674 (2000).
\bibitem{PMMA-CILIBERTO} L. Bellon, S. Ciliberto, C. Laroche, Eur. Phys. J. B {\bf 25,} 223 (2002).
\bibitem{PMMA}  K. Fukao and A. Sakamoto, cond-mat/0410602.
\bibitem{AGINGCMR} P. Levy, F. Parisi, L. Granja, E. Indelicato and
G. Polla, Phys. Rev. Lett. {\bf 89,} 137001 (2002).
\bibitem{TWINS} P.E. J\"onson, H. Yoshino and P. Nordblad, Phys. Rev. Lett.
{\bf 89,} 97201 (2002).
\bibitem{GHOST} P.E. J\"onsson, R. Mathieu, P. Nordblad, H. Yoshino,
H. Aruga Katori and A. Ito, Phys. Rev. B {\bf 70,} 174402 (2004).
\bibitem{COHE-LENGTH} F. Bert, V. Dupuis, E. Vincent, J. Hammann and
  J.-P. Bouchaud, Phys. Rev. Lett. {\bf 92,} 167203 (2004).
\bibitem{ENERGYPICT} J.-P. Bouchaud and D.S. Dean, J. Phys. I (France)
{\bf 5,} 265 (1995); M. Sales, J.-P. Bouchaud and F. Ritort,
J. Phys. A: Math. Gen. {\bf 36,} 665 (2003); M. Sasaki, V. Dupuis,
J.-P. Bouchaud and E. Vincent, Eur. Phys. J. B {\bf 29,} 469 (2002).
\bibitem{BOUCHAUD} J.P. Bouchaud, {\em Soft and Fragile matter}, Eds:
M.E. Cates, M.R. Evans (Institute of Physics Publishing, 2000).
\bibitem{YOSHINO} T. Komori, H, Yoshino,
H. Takayama, J. Phys. Soc. Jpn. {\bf 69,} Suppl., 228 (2000).
\bibitem{RICCI} M. Picco, F. Ricci-Tersenghi, F. Ritort,
Phys. Rev. B {\bf 63}, 174412 (2001).
\bibitem{BERTHIER} L. Berthier, J.-P. Bouchaud, Phys. Rev. B {\bf 66},
054404 (2002).
\bibitem{ROMANOS} A. Maiorano, E. Marinari and F. Ricci-Tersenghi,
cond-mat/0409577.
\bibitem{BERNARDI} See L.W. Bernardi, H. Yoshino, K. Hukushima, H. Takayama, A. Tobo and A. Ito, Phys. Rev. Lett. {\bf 86,} 720 (2001), and references there in. 
\bibitem{RICCI-PRIVATE} F. Ricci-Tersenghi, private communication.
\bibitem{SASAKI} M. Sasaki, O.C. Martin, Phys. Rev. Lett. {\bf 91,}
097201 (2003).
\bibitem{C4DEF} E. Marinari, G. Parisi, F. Ricci-Tersenghi and
J.J. Ruiz-Lorenzo, J. Phys. A {\bf 33},2373 (2000).
\bibitem{JAP-GROWTH} T. Komori, H. Yoshino and H. Takayama,
J. Phys. Soc. Jpn. {\bf 68,} 3387 (1999).
\bibitem{DILUTED} H. G. Ballesteros, L. A. Fern\'andez, V. Mart\'{\i}n-Mayor,
A. Muñoz Sudupe, G. Parisi and J. J. Ruiz-Lorenzo, Phys. Rev. B, {\bf
58} 2740 (1998).
\bibitem{SUE} A. Cruz, J. Pech, A. Taranc\'on, P. T\'ellez,
C. L. Ullod and C. Ungil, Comput. Phys. Commun. {\bf 133,} 165 (2001).
\bibitem{SERGIOTESIS} S. Jim\'enez, Ph.D. Thesis, U. Zaragoza, January
2005.
\bibitem{CHAOS} A.J. Bray, M.A. Moore,
Phys. Rev. Lett. {\bf 58}, 57 (1987)
\bibitem{FDT} L. F. Cugliandolo, J. Kurchan Phys. Rev. Lett. {\bf 71,}
173 (1993); S. Franz, H. Rieger, J. Stat. Phys. {\bf 79} 749 (1995);
E. Marinari, G. Parisi, F. Ricci-Tersenghi and J.J. Ruiz-Lorenzo,
J. Phys. A: Math. Gen. {\bf 31}, 2611 (1998); D. H\'erisson, M. Ocio,
Phys. Rev. Lett. {\bf 88}, 257202 (2002).
\bibitem{RIZZO} T. Rizzo and A. Crisanti, Phys. Rev. Lett. {\bf 90,}
137201 (2003).
\bibitem{COOPER} F. Cooper, B. Freedman, D. Preston,
Nucl. Phys. {\bf B210}, 210 (1989).
\bibitem{SUEFSS} H.G. Ballesteros, A. Cruz, L. A. Fern\'andez,
V. Mart\'{\i}n-Mayor, J. Pech, J. J. Ruiz-Lorenzo, A. Taranc\'on,
P. T\'ellez, C. L. Ullod, and C. Ungil, Phys. Rev. B {\bf 62,} 14237
(2000).
\bibitem{BOUCHAUDPRIVATE} Parametrizations of $C_4(r,t)$ different
from Eq.(\ref{ANSATZ}) can be found, where $\xi$ does not decrease
(J.P. Bouchaud and L. Berthier, private communication).

\bibitem{SUEAGING} S. Jim\'enez, V. Mart\'{\i}n-Mayor, G. Parisi and
A. Taranc\'on, J. Phys. A {\bf 36,} 10755 (2003).

\bibitem{RIEGER} H. Rieger, J. Phys. A{\bf 26}, L615 (1993);
J. Kisker, L. Santen, M. Schreckenberg and H. Rieger, Phys. Rev. B
{\bf 53,} 6418 (1996).
\bibitem{BERTHIERYOUNG} L. Berthier and A.P. Young, preprint cond-mat/0503012.
\bibitem{FLOREN} F. Krzakala, Europhys. Lett., {\bf 66,} 847 (2004).
\bibitem{HUKUSHIMA-IBA} Hukushima and Iba,  preprint cond-mat/0207123.
\bibitem{BERTHIER2003} L. Berthier and J.-P. Bouchaud,
Phys. Rev. Lett. {\bf 90,} 059701 (2003).
\end{thebibliography}
\end{document}